\documentclass[a4paper,11pt]{article} \usepackage{pos}

\usepackage{caption}
\usepackage[skip=2pt]{subcaption}
\usepackage{microtype}

\usepackage{titlesec}

\titlespacing\section{0pt}{13pt plus 6pt minus 2pt}{5pt plus 4pt minus 2pt}
\titlespacing\subsection{0pt}{11pt plus 6pt minus 2pt}{5pt plus 4pt minus 2pt}
\titlespacing\paragraph{0pt}{6pt plus 6pt minus 4pt}{0pt plus 4pt minus 2pt}[4pt]

\let\OLDthebibliography\thebibliography
\renewcommand\thebibliography[1]{
  \OLDthebibliography{#1}
  \setlength{\parskip}{0pt}
  \setlength{\itemsep}{0pt plus 0.3ex}
}

\usepackage{grffile}
\usepackage{slashed}
\usepackage{xspace}

\title{Towards the $\beta$ function of $\su{2}$ with adjoint matter using Pauli--Villars fields}
\author*[a]{Ed Bennett}
\affiliation[a]{Swansea Academy of Advanced Computing, Swansea University, Bay
  Campus, Fabian Way, Swansea SA1 8EN, United Kingdom}
\emailAdd{e.j.bennett@swansea.ac.uk}

\author[b]{Andreas Athenodorou}
\affiliation[b]{Computation-based Science and Technology Research Center, The Cyprus Institute, 20 Kavafi Str., Nicosia 2121, Cyprus}
\emailAdd{a.athenodorou@cyi.ac.cy}

\author[c]{Georg Bergner}
\affiliation[c]{University of Jena, Institute for Theoretical Physics, Max-Wien-Platz 1, D-07743 Jena, Germany}
\emailAdd{georg.bergner@uni-jena.de}

\author[d,e]{Pietro Butti}
\affiliation[d]{Departamento de Física Teórica, Facultad de Ciencias and Centro de Astropartículas y Física de Altas Energías (CAPA),
  Universidad de Zaragoza, Calle Pedro Cerbuna 12, E-50009, Zaragoza, Spain}
\affiliation[e]{Quantum Field Theory Center ($\hbar$QTC) at IMADA \& D-IAS, Southern Denmark University, Campusvej 55, 5230 Odense M, Denmark}
\emailAdd{pbutti@qtc.sdu.dk}

\author[a,f]{Biagio Lucini}
\affiliation[f]{Department of Mathematics, Swansea University, Fabian Way, Swansea SA1 8EN, UK}
\emailAdd{b.lucini@swansea.ac.uk}

\abstract{
  The family of SU(2) theories with matter transforming in the adjoint representation
  has attracted interest from many angles.
  The two-flavour theory, known as Minimal Walking Technicolor,
  has a body of evidence pointing to
  it being in the conformal window with anomalous dimension $\gamma_{*}\approx0.3$.
  Perturbative calculations would suggest that the one-flavour theory should be confining and chirally broken;
  however,
  lattice studies of the theory have been inconclusive.
  In this contribution we present a first look at efforts towards the computation of the beta function of these theories using the gradient flow methodology.
  Following an exploration of the phase diagram of the two theories with Wilson fermions and additional Pauli–Villars fields,
  we tune the bare fermion mass to near the chiral limit,
  and subsequently generate ensembles at five lattice volumes and a range of lattice spacings.
}

\FullConference{The 41st International Symposium on Lattice Field Theory (LATTICE2024)\\
 28 July - 3 August 2024\\
Liverpool, UK\\}

\DeclareMathOperator{\Tr}{Tr}

\newcommand\Sgroup[2]{\mathrm{#1}(#2)\xspace}
\newcommand\su[1]{\ensuremath{\Sgroup{SU}{#1}}}

\newcommand\Nx[1]{\ensuremath{N_{\textnormal{#1}}}\xspace}
\newcommand\Nf{\Nx{f}}

\newcommand\NPV{\Nx{PV}}

\newcommand\mx[1]{\ensuremath{m_{\textnormal{#1}}}\xspace}
\newcommand\mPV{\mx{PV}}
\newcommand\mf{\mx{f}}
\newcommand\mpcac{\mx{PCAC}}

\newcommand\gGF{g_{\mathrm{GF}}}
\newcommand\dee{\mathrm{d}}

\begin{document}
\maketitle

\section{Introduction}

In the last two decades
the $\su{2}$ gauge theories with various numbers $\Nf$ of adjoint fermion flavours
have been studied extensively on the lattice.
This family of theories is interesting in a number of contexts:
the $\Nf=2$ theory, for example,
provides the minimal possible realisation of Walking Technicolor.
Numerous studies have placed it in the conformal window;
recent work~\cite{Athenodorou:2024rba} shows
a mass anomalous dimension of 0.304(4) at finite lattice spacing
from the scaling of the Dirac mode number,
while earlier step-scaling studies~\cite{Rantaharju:2015cne} indicate
an anomalous dimension of $0.263(4){}^{+0.012}_{-0.015}$.
The $\Nf=1$ theory cannot by itself break electroweak symmetry,
although it can be extended in such a direction in theories such as Ultra Minimal Walking Technicolor,
and is interesting in the context of being able to
identify the lower end of the conformal window,
and to observe near-conformal behaviour.
It is also relevant to studies of topological phase transitions
via 't Hooft anomaly matching.
Recent work~\cite{Athenodorou:2024rba} finds
a relatively small continuum limit mass anomalous dimension of 0.170(7)
by extrapolating values computed using the mode number at finite lattice spacing.

The fixed point in the $\Nf=2$ theory,
and any possible remnant of a fixed point in the $\Nf=1$ theory,
can only be found at relatively strong coupling.
Such couplings are challenging to reach on the lattice,
as in this regime the lattice spacing becomes large,
and a bulk phase transition can mean that results are not connected to the continuum limit.
Recent work~\cite{Hasenfratz:2021zsl,Hasenfratz:2024fad}
has suggested that introducing Pauli--Villars (PV) fields into the action
may reduce lattice artefacts at strong coupling
and hence widen the range of couplings at which a theory may be studied.

\section{Summary of the theory}

We generate ensembles using the Wilson gauge action
$S_{\mathrm{G}} = \beta_{0} \sum_{p} \Tr \left[ 1 - \frac{1}{2} U(p) \right]$
and Wilson fermion action,
% \begin{align}
%   S_{\mathrm{G}} &= \beta_{0} \sum_{p} \Tr \left[ 1 - \frac{1}{2} U(p) \right]\;,
%   & S_{\mathrm{F}} &= \sum_{\alpha=1}^{\Nf} \overline{\psi}_{\alpha} (x)(i\slashed{D}-m)\psi_{\alpha}(x)\;,
% \end{align}
$S_{\mathrm{F}} = \sum_{\alpha=1}^{\Nf} \overline{\psi}_{\alpha} (x)(i\slashed{D}-m)\psi_{\alpha}(x)$,
where $\Nf\in\{1,2\}$,
as in our previous work~\cite{Athenodorou:2024rba}.
To this,
we add \NPV species of PV field each with mass \mPV,
implemented as a Hasenbusch mass term
with a very heavy mass
% ($\mf = 10.0$)
in the denominator.

While previous work~\cite{Hasenfratz:2021zsl,Hasenfratz:2024fad}
has smeared these PV fields,
in this work we use unsmeared fields.
Theoretically,
since this has the effect of adding a single plaquette contribution to the action,
this would be expected to have no effect on the lattice artefacts,
leading mainly to a shift in the value of $\beta_{0}$.
In this work we will test that prediction.

We make use of the RHMC and HMC algorithms respectively for $\Nf=1$ and $2$.
The HiRep~\cite{DelDebbio:2008zf,hirep-github} code is used in both cases,
and is run both on CPU
(for the phase diagram)
and on A100 GPUs
(for larger volumes).
GNU Parallel~\cite{tange_2024_11043435}
is used to manage large swarms of small jobs.
Meson correlation function and gradient flow histories
are computed using HiRep on CPUs.

\section{Parameter tuning}

\begin{figure}
  \begin{subfigure}{\linewidth}
    \includegraphics[width=\textwidth]{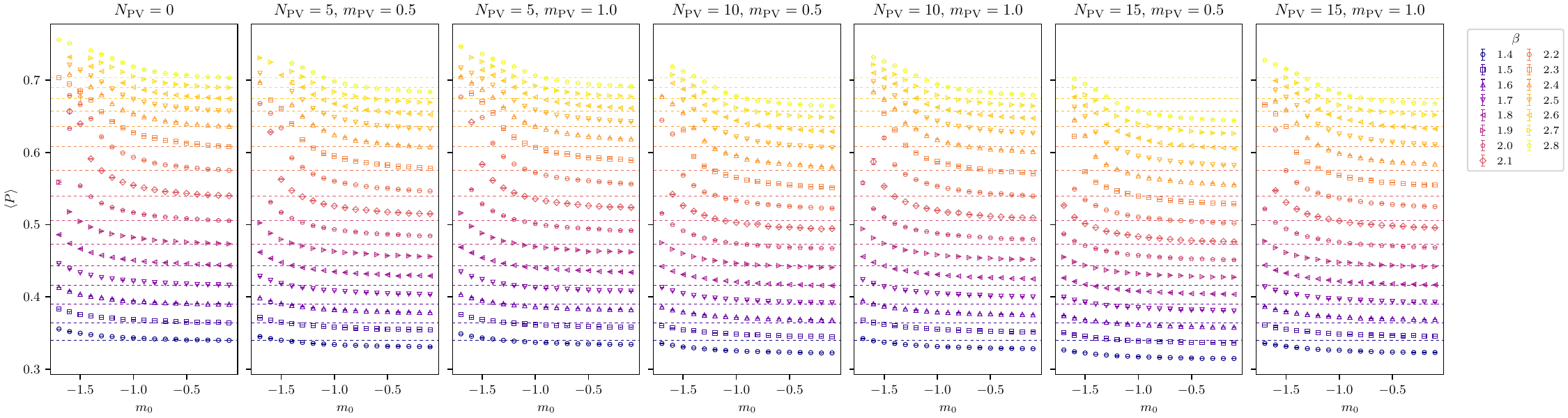}
    \caption{$\Nf=1$}
  \end{subfigure}

  \begin{subfigure}{\linewidth}
    \includegraphics[width=\textwidth]{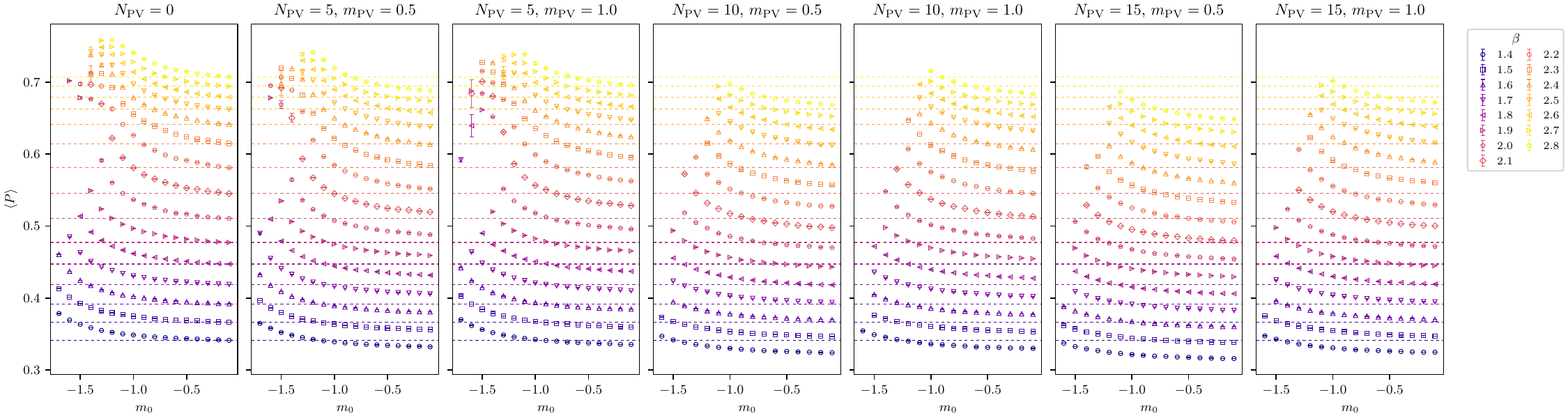}
    \caption{$\Nf=2$}
  \end{subfigure}

  \caption{\label{fig:phasediagram}Phase diagram of the theory
    with (a) $\Nf=1$ and (b) $\Nf=2$ flavours of Dirac fermion
    in the adjoint representation,
    with $\NPV \in \{0,5,10,15\}$ and $\mPV\in\{0.5, 1.0\}$
    (panels left to right).
    The average plaquette is plotted as a function of the bare fermion mass $\mf\in[-1.7, -0.1]$ (horizontal axis)
    and the bare coupling $\beta_{0} \in [1.8, 2.8]$ (colours).
    For each value of $\beta_{0}$,
    the value of the heaviest mass $m=-0.1$ is projected across all panels
    as dashed lines of the same colour,
    to allow easier comparison.
  }
\end{figure}

Since we expect the PV fields to affect the region of parameter space accessible to computation,
we must map out that parameter space to identify the region to study.
%To this end,
We study the average plaquette on an $8^{4}$ lattice as a function of $\beta_{0}$ and $\mf$,
for both $\Nf=1$ and $2$,
with $\NPV=5$, $10$, and $15$,
each with $\mPV=0.5$ and $1.0$.
We also include the case $\NPV=0$ for comparison.
% in both cases.
% at both values of \Nf.
We show the resulting phase diagrams in Fig.~\ref{fig:phasediagram};
the shift in $\beta_{0}$ is clearly visible here.
Based on this,
we identify regions of $\beta_{0}$ to target for
$\Nf=1$, $\NPV\in\{5,10,15\}$, $\mPV=0.5$
and $\Nf=2$, $\NPV=15$, $\mPV=0.5$.

\begin{figure}
  \begin{subfigure}{0.495\textwidth}
    \center
    \includegraphics[width=0.8\textwidth]{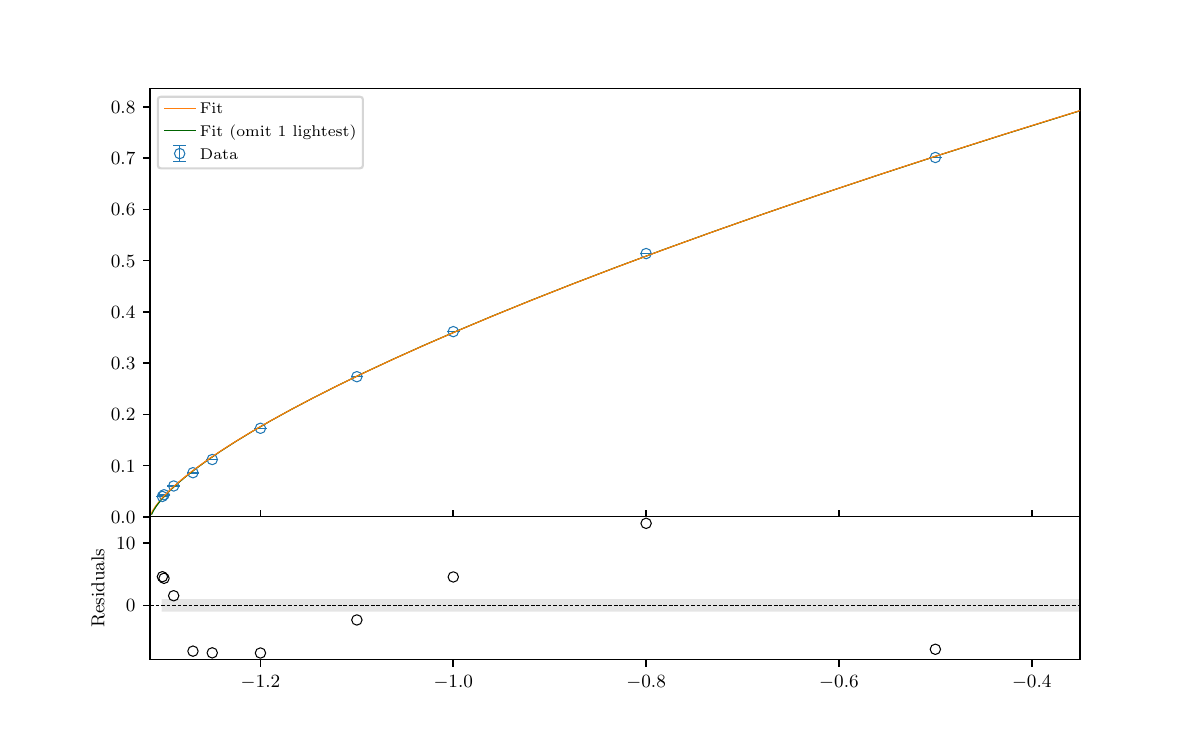}
    \caption{$\Nf=1$, $\NPV=15$, $\mPV=0.5$, $\beta=2.7$}
  \end{subfigure}
  \begin{subfigure}{0.495\textwidth}
    \center
    \includegraphics[width=0.8\textwidth]{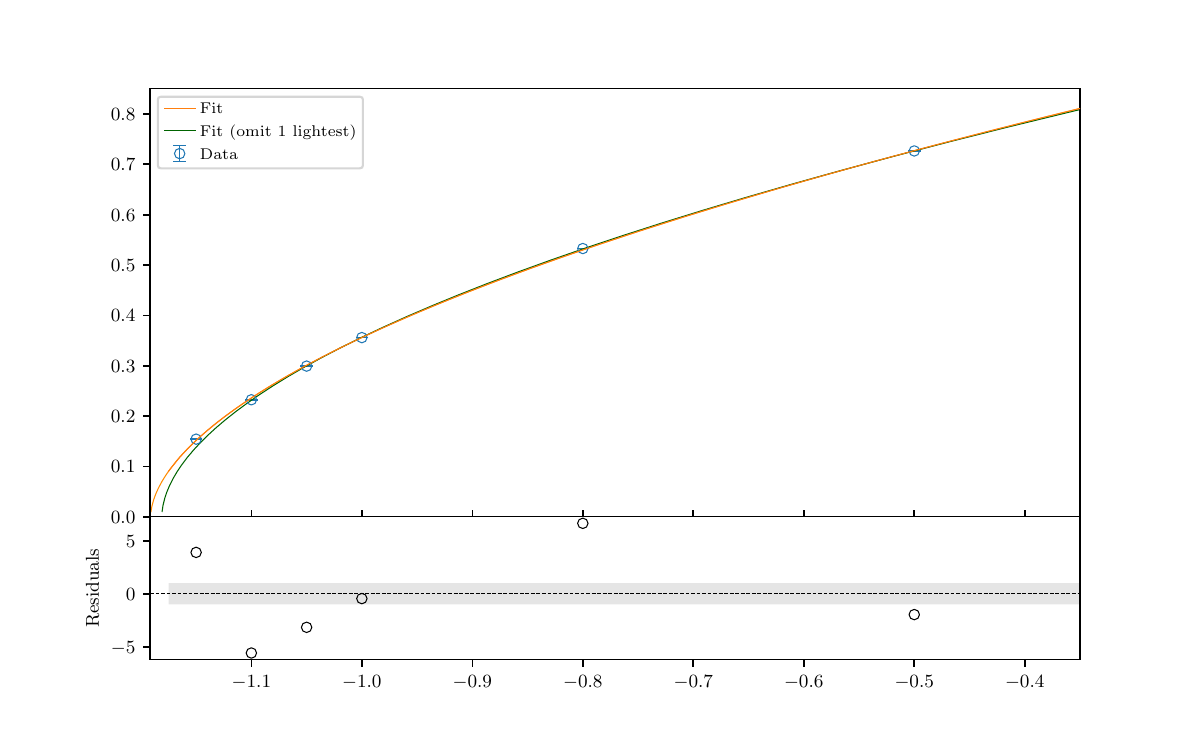}
    \caption{$\Nf=2, \NPV=5, \mPV=0.5, \beta=2.35$}
  \end{subfigure}
  \caption{\label{fig:mass-extrapolation}
    The PCAC fermion mass as a function of the bare fermion mass
    for two specific choices of $\Nf$, $\NPV$, $mPV$, $\beta$.
    In addition to the data,
    the fit line and residual at each point are also shown.
  }
\end{figure}

To compute the $\beta$ function,
we aim to simulate at or very close to the chiral limit.
Because the Wilson fermion action introduces
an additive renormalisation to the fermion mass,
it is necessary to tune the bare fermion mass $m_{0}$
to find the chiral limit
for each value of $\NPV$, $\mPV$, $\beta_{0}$ considered.
To do this,
in each case,
a range of ensembles are generated on a $32 \times 16^{3}$ volume,
and the PCAC mass~\cite{DelDebbio:2007wk} is computed.
%as a renormalised fermion mass.
This is then extrapolated to the chiral limit using
$\mpcac = B(m_{0} - m_{0}^{\mathrm{cr}})^{C}$,
where $B$, $C$, and $m_{0}^{\mathrm{cr}}$ are fit parameters,
the latter of which is the chiral limit bare mass
that is used for full ensemble generation.
Examples of this extrapolation are shown in Fig.~\ref{fig:mass-extrapolation}.

\section{Computing the running coupling and $\beta$ function}

In this section we will summarise the methodology of computing
the running coupling and $\beta$ function
described by the authors of Ref.~\cite{Hasenfratz:2024fad},
to which we refer for further detail.

We may evolve our field configurations under the diffusion equation
defined by the gradient of the Wilson action
(the Wilson flow),
as described in Ref.~\cite{Luscher:2010iy}.
The running coupling at finite volume and lattice spacing
can be computed as a function of the flow time $t$ as
%\begin{equation}
$\gGF^{2}(t;L, g_{0}^{2}) = \langle t^{2} E(t) \rangle$,
%\end{equation}
where the energy density $E$ may be computed on the lattice via a variety of operators;
in this work we consider the plaquette and clover operators.
From this,
the $\beta$ function may be computed as
%\begin{equation}
  $\beta(t; L, g_{0^{2}}) = t \frac{\dee}{\dee t}\gGF^{2}(t; L, g_{0}^{2})$,
%\end{equation}
where we may fit subsets of the volumes considered,
and compute the fit results' mean weighted using a modified
Akaike Information Criteria.

Both $\gGF^{2}$ and $\beta$ may be extrapolated to the infinite-volume limit
using a linear fit form in the reciprocal of the lattice volume
\begin{align}
  \gGF^{2}(t; L, g_{0}^{2}) &= \gGF^{2}(t; g_{0}^{2})+C_{g^{2}}L^{-4}\;, &
  \beta(t; L, g_{0}^{2}) &= \beta(t; g_{0}^{2}) + C_{\beta}L^{-4}\;.\label{eq:infinite-volume}
\end{align}
These may then be interpolated against each other to obtain
the $\beta$ function at arbitrary $\gGF^{2}$,
using
\begin{equation}
  \beta^{\mathrm{int}}(t, \gGF^{2}) = \gGF^{4}\sum_{n=0}^{N-1}p_{n}\gGF^{2n}\;,\label{eq:beta-interp}
\end{equation}
where $N$ is fixed to the lowest value that gives a reasonable fit.
This allows the $\beta$ function to be scanned across a range of running couplings,
and at each of these a continuum limit may be taken by extrapolating the linear Ansatz
\begin{equation}
  \beta^{\mathrm{int}}(t, \gGF^{2}) = \beta^{\mathrm{cont}}(\gGF^{2})+\frac{C_{\mathrm{cont}}}{t}\;.
\end{equation}

\section{Analysis workflow}

The data analysis pipeline is written using Snakemake~\cite{molder2021sustainable},
connecting individual tools each of which makes use of pyerrors~\cite{Joswig:2022qfe}.
This allows simple rules
(for example,
how to transform one or more input data files
to an output containing one or more statistical quantities)
to be easily composed,
independent steps to be parallelised,
and redundant computations to be skipped.
To generate every plot in this contribution,
excluding Fig.~\ref{fig:hp-nf12},
starting from the raw output files transferred from the HPC resources used to generate them,
requires the single command:
% \\
% \hphantom{indent}\texttt{
\begin{verbatim}
snakemake --cores 6 --use-conda
\end{verbatim}
%} \\to
be run twice,
once per \Nf.
This then launches 344 (for $\Nf=1$) + 81 (for $\Nf=2$) job steps,
taking 5 minutes to run on six cores of an Apple M1 Pro CPU.
All required software packages are installed automatically;
no setup is required beyond installing Snakemake.

\begin{figure}
  \center
  \includegraphics[width=0.8\textwidth]{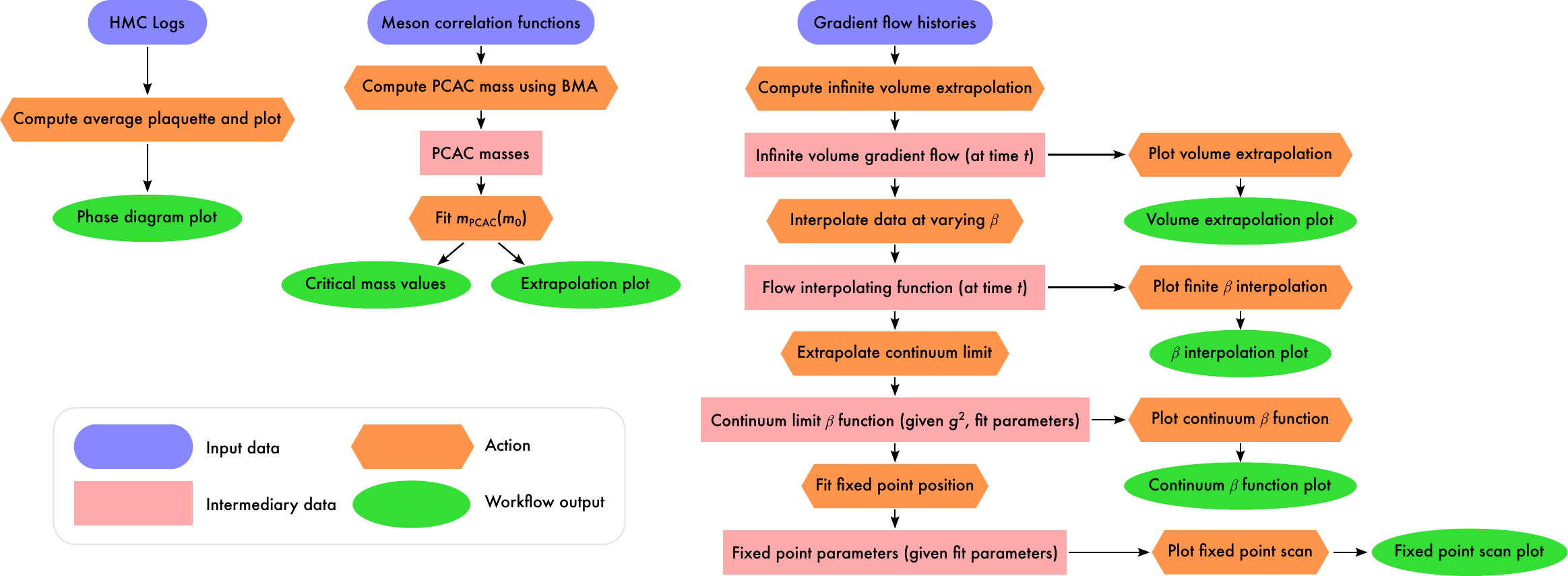}

  \caption{\label{fig:flowchart}
    Diagram of the analysis workflow used.
    Blue lozenges represent input data,
    generated using HPC facilities and copied off for analysis.
    Pink rectangles represent intermediary data files,
    which are automatically generated and updated as necessary by the workflow.
    Orange hexagons represent specific rules, transforming input to output files.
    Green ellipses represent output files,
    either data or plots.
    Each may have multiple instances,
    for different input parameters.
    (For example,
    there is more than one extrapolation plot of \mpcac produced,
    each relies on more than one \mpcac value,
    each requiring an execution of the effective mass fit.)
  }
\end{figure}

An illustration of the workflow is shown in Fig.~\ref{fig:flowchart};
each rule
(orange hexagon)
may be run many times with different inputs to produce different output files.
If only one input file has changed since the previous execution,
then only the steps that directly or indirectly depend on that input are re-run.

\begin{figure}
  \center
  \includegraphics{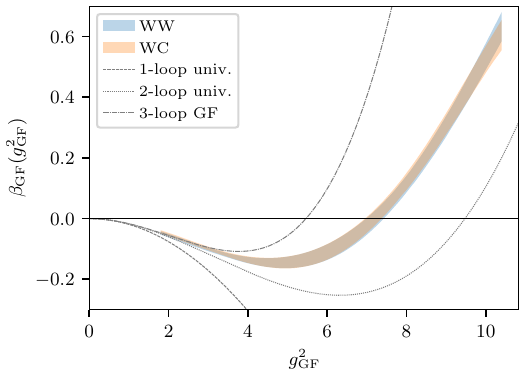}

  \caption{\label{fig:hp-nf12}
    The continuum $\beta$ function for the \su{3} theory with $\Nf=12$ fundamental flavours,
    computed using the dataset at Ref.~\cite{hp-data}.
    This qualitatively agrees with Fig.~1 of Ref.~\cite{Hasenfratz:2024fad},
    which presented the first analysis of the same data.
  }
\end{figure}

In order to verify that the implementation of the analysis was correct,
a modified version of the workflow was applied to the open dataset~\cite{hp-data}
used to prepare Ref.~\cite{Hasenfratz:2024fad}.
The workflow reproduces the finding of a conformal fixed point
in the $\beta$ function of the \su{3} theory with 12 fundamental Dirac flavours.
Reproducing every figure\footnote{
  Excluding Fig.~2,
  the data for which were not released,
  and Fig.~9,
  which is not directly relevant to this work.
} of Ref.~\cite{Hasenfratz:2024fad}
% starting
from the released data
requires the same single command as above,
which launches 4,227 job steps,
and takes 19 minutes on six cores of an Apple M1 Pro CPU.
The results,
such as the continuum $\beta$-function shown in Fig.~\ref{fig:hp-nf12}
are qualitatively consistent;
small differences in the numerical results
(for example,
finding the fixed point at $\gGF^{2}\approx 7.0$ rather than $6.5$)
are likely due to subtly different weights in some
% of the
extrapolations.

\section{Results}

In this section,
we present preliminary results for the gradient flow $\beta$ function
of \su{2} with $\Nf=2$ flavours of adjoint Dirac fermion.
The $\Nf=1$ theory,
relying on the slower RHMC algorithm,
has not produced sufficient data to reliably make an infinite volume extrapolation.

In all cases we generate data at the five volumes $L^{4}$,
$L \in \{24, 28, 32, 36, 40\}$,
with a molecular dynamics trajectory length $t_{\mathrm{len}}\in[0.4, 2.0]$.
In all cases we cut the first 2000 molecular dynamics time units (MDTU) for thermalisation,
and we aim generate at least 6000 MDTU of thermalised data.
However,
some points in this contribution have fewer statistics than this,
as ensembles were still being generated at the time of writing.

%\subsection{Infinite volume extrapolation}

\begin{figure}
  \center
  \includegraphics[width=\textwidth]{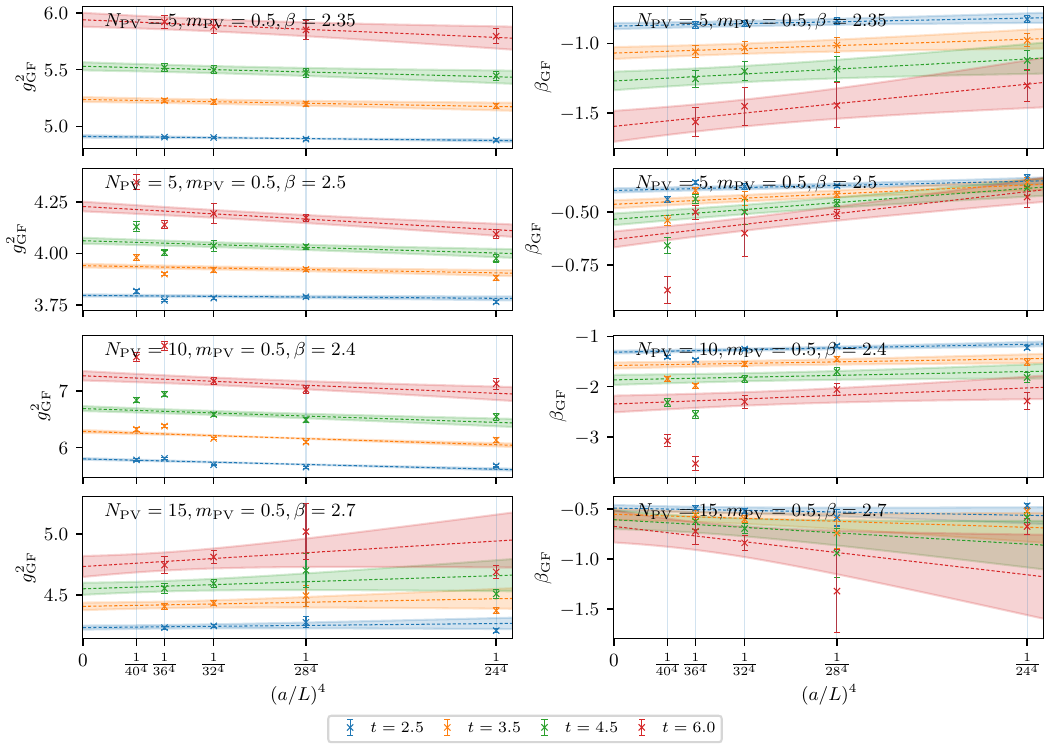}

  \caption{\label{fig:infinite-volume}
    The finite-volume data (points)
    and infinite volume extrapolations using Eq.~\eqref{eq:infinite-volume} (bands)
    of the running coupling (left)
    and the $\beta$ function (right)
    of the theory with $\Nf=2$
    at four values of the flow time $t\in\{2.5, 3.5, 4.5, 6.0\}$,
    for the four cases $(\NPV=5, \beta=2.35)$, $(\NPV=5, \beta=2.5)$, $(\NPV=10, \beta=2.4)$, and $\NPV=15, \beta=2.7$. $\mPV=0.5$ in all cases.
    Blue vertical lines indicate the volumes studied in this work;
    where points are missing,
    the relevant data have not yet generated enough statistics to analyse.
  }
\end{figure}

We extrapolate $\gGF^{2}$ and $\beta$
in the $\Nf=2$ theory
to the infinite volume limit
using Eq.~\eqref{eq:infinite-volume}.
Selected extrapolations are shown in Fig.~\ref{fig:infinite-volume}.
There remains some visible non-linear volume dependence at relatively large volume;
this may be due to underestimated statistical uncertainties
from low-statistics ensembles that currently lack sufficient ergodicity
to give a good estimate of the statistical error.

%\subsection{Spanning of parameter space}

\begin{figure}
  \center
  \includegraphics[width=0.8\textwidth]{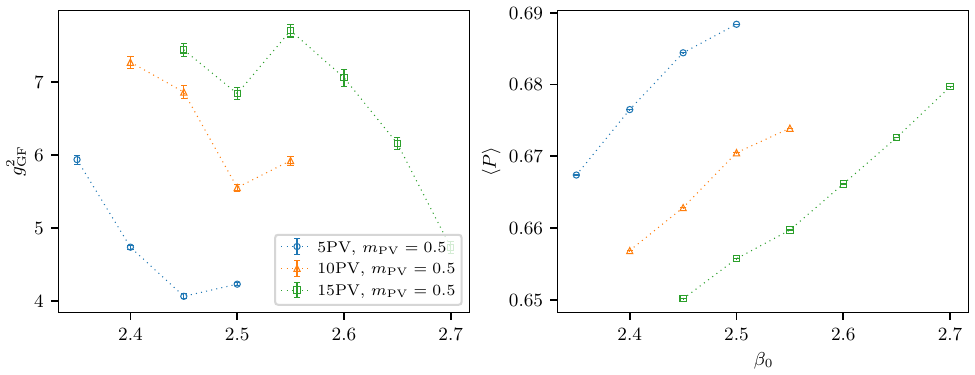}

  \caption{\label{fig:parameter-span}
    Plots of the running coupling and average plaquette
    as a function of the bare gauge coupling $\beta_{0}$,
    for $\Nf=2$, $\NPV\in\{5,10,15\}$, $\mPV=0.5$.
    The three cases span a relatively similar range of $\gGF^{2}$ and $\langle P \rangle$.
  }
\end{figure}

In Fig.~\ref{fig:parameter-span} we show how
the range of running coupling and average plaquette spanned
changes with $\NPV$.
In contrast to previous work with smeared PV fields~\cite{Hasenfratz:2021zsl},
where larger $\NPV$ significantly raised $\langle P \rangle$,
and allowed a larger range of $\gGF^{2}$ to be spanned,
in this case there is no significant shift in $\gGF^{2}$ or $\langle P \rangle$,
consistent with our expectations for unsmeared PV fields.

%\subsection{Interpolating at finite running coupling}

\begin{figure}
  \center
  \includegraphics[width=0.8\textwidth]{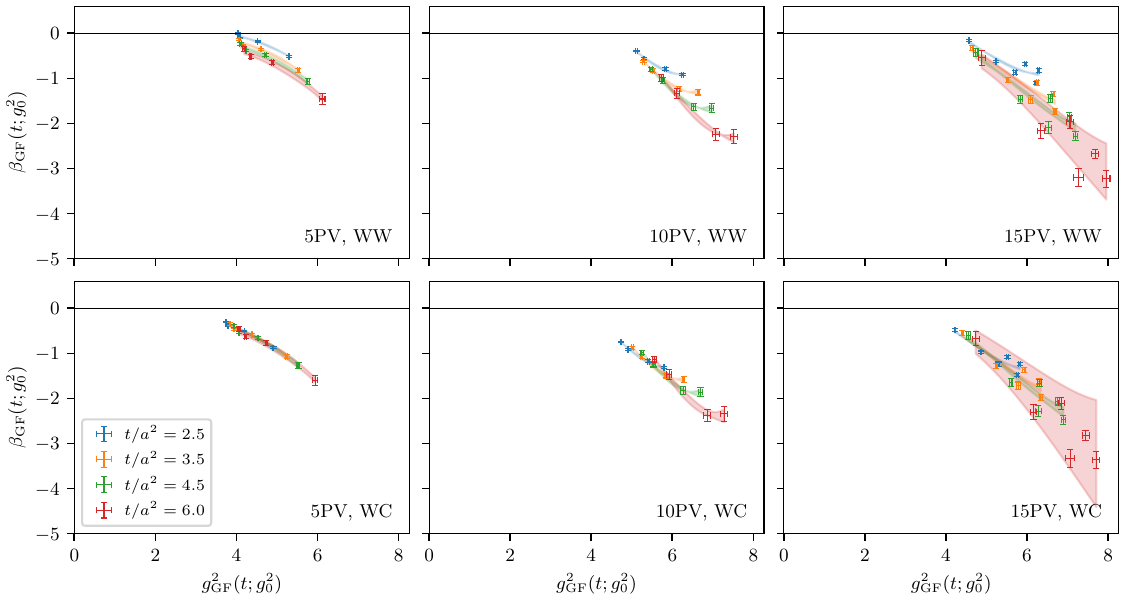}

  \caption{\label{fig:interpolating-g2}
    Plots of the gradient flow $\beta$ function of
    \su{2} with $\Nf=2$ flavours of adjoint Dirac fermion,
    with $\NPV\in\{5,10,15\}$ PV fields (left to right),
    computed using the Wilson plaquette (upper) and symmetric clover (lower) operators.
  }
\end{figure}

We may interpolate our data for $\beta$ and $\gGF^{2}$ using Eq.~\eqref{eq:beta-interp};
this is a necessary step to eventually be able to take the continuum limit,
to vary $t$ at constant $\gGF^{2}$.
We show these interpolations in Fig.~\ref{fig:interpolating-g2}.
We observe that the data computed using the clover operator
are significantly more consistent as $t/a^{2}$ increases
(and hence the continuum is approached)
than those computed with the plaquette operator;
this matches our expectations that the clover operator should have fewer finite lattice spacing effects.
The relatively narrow range of $\gGF^{2}$ spanned,
and the shift in $\gGF^{2}$ as $t/a^{2}$ increases,
means that the range of $\gGF^{2}$
that could be interpolated at a wide range of $t/a^{2}$
is not sufficient to estimate the continuum $\beta$ function
for a useful range of $\gGF^{2}$.

\section{Conclusions}

We have computed the gradient flow beta function
for \su{2} with $\Nf=2$ flavours of adjoint Dirac fermion,
with the addition of $\NPV=5$, 10, and 15 unsmeared PV fields.
We have also studied the parameter space of the equivalent $\Nf=1$ theory.
Unlike results seen in work performed with smeared PV fields~\cite{Hasenfratz:2024fad},
we do not see a substantial change in the phase diagram
or the range of running coupling that may be observed
as \NPV is increased,
instead only seeing a constant shift in the phase diagram as a function of the bare coupling;
this is consistent with theoretical expectations.

We now aim to repeat this work with smeared PV fields,
such that we may compute the $\beta$ function at stronger coupling,
including observing the fixed point in the $\Nf=2$ case,
and any remnant of a fixed point in the $\Nf=1$ case.

\section*{Acknowledgements}

We would like to thank Anna Hasenfratz for valuable conversations.

A.A.\ was supported by the Horizon 2020 European research infrastructures
programme ``NI4OS-Europe” with grant agreement no.\ 857645, by ``SimEA" project funded by the European Union’s Horizon 2020 research and innovation programme under grant agreement No 810660, as well as by the ``EuroCC" project funded by the ``Deputy Ministry of Research, Innovation and Digital Policy and the Cyprus Research and Innovation Foundation" as well as by the EuroHPC JU under grant agreement No.~101101903.
The work of E.B.\ has been supported by the STFC Research Software Engineering Fellowship EP/V052489/1. The work of E.B.\ and B.L.\ has been supported in part by the EPSRC ExCALIBUR programme ExaTEPP (project EP/X017168/1) and the STFC Consolidated Grant No. ST/T000813/1.
G.B.\ is funded by the Deutsche Forschungsgemeinschaft (DFG) under Grant No.~432299911 and 431842497.
P.B.\ and B.L.\ acknowledge support by the project H2020-MSCAITN-2018-813942 (EuroPLEx) and the EU Horizon 2020 research and innovation programme. P.B.\ additionally acknowledges support by the Grant DGA-FSE grant 2020-E21-17R Aragon Government and the European Union - NextGenerationEU Recovery and Resilience Program on ``Astrofísica y Física de Altas Energías" CEFCA-CAPA-ITAINNOVA.

This work used the DiRAC Extreme Scaling service Tursa
at the University of Edinburgh,
managed by EPCC
on behalf of the STFC DiRAC HPC Facility
(www.dirac.ac.uk).
The DiRAC service at Edinburgh was funded by
BEIS,
UKRI
and STFC capital funding and STFC operations grants.
DiRAC is part of the UKRI Digital Research Infrastructure.
We acknowledge the support of the Supercomputing Wales project,
which is part-funded by the European Regional Development Fund
(ERDF)
via Welsh Government.
The data from Ref.~\cite{hp-data} were generated
using the computing and long-term storage facilities of the USQCD Collaboration,
which are funded by the Office of Science of the U.S. Department of Energy,
and the Alpine high performance computing resource at the University of Colorado Boulder.
Alpine is jointly funded by the University of Colorado Boulder,
the University of Colorado Anschutz,
and Colorado State University.

\paragraph*{Open access statement}
\hphantom{gap}For the purpose of open access, the authors have applied a Creative Commons
Attribution (CC BY) licence to any Author Accepted Manuscript version arising.
The original image of
the poster on which this contribution is based
is available at Ref.~\cite{poster}
under the same license.

\paragraph*{Research Data Access Statement}
\hphantom{gap}The data generated for this manuscript can be downloaded from Ref.~\cite{datapackage},
and the workflow used to analyse it from Ref.~\cite{analysiscode}.
The analysis workflow used to analyse the open data at Ref.~\cite{hp-data}
is available from Ref.~\cite{hp-analysis}.

\small
\bibliography{references} \bibliographystyle{apsrev}
\end{document}